\documentclass[11pt]{article}
\usepackage{times}
\usepackage{amsmath}
\usepackage{arxiv}
\PassOptionsToPackage{hyphens}{url}\usepackage[hyperref]{acl2019}
\aclfinalcopy
\usepackage{tabularx}
\usepackage[utf8]{inputenc} % allow utf-8 input
\usepackage{cuted}
\usepackage{longtable}
\usepackage{adjustbox}
\usepackage{hyperref}       % hyperlinks
\usepackage{url}            % simple URL typesetting
\usepackage{booktabs}       % professional-quality tables
\usepackage{amsfonts}       % blackboard math symbols
\usepackage{nicefrac}       % compact symbols for 1/2, etc.
\usepackage{lipsum}
\usepackage{graphicx}
\graphicspath{ {./images/} }
\usepackage[toc,page]{appendix}
\usepackage{graphicx}
\graphicspath{ {./images/} }
\usepackage{array}
\usepackage{supertabular,booktabs}

\title{Content Aware Analysis of Scholarly Networks:\\ A Case Study on CORD19 Dataset}
\usepackage{fancyhdr}       % header
\usepackage{booktabs}       % professional-quality tables

\author{
 Mehmet Emre Akbulut$^*$\\
  Politecnico di Milano\\
  \texttt{mehmetemre.akbulut@mail.polimi.it} \\
     \And
 Yusuf Erdem Nacar\thanks{~~Equal contribution.}$^*$\\
  Technical University of Munich\\
  \texttt{yusuferdem.nacar@tum.de} \\
}

\date{}
\begin{document}

\maketitle
\begin{abstract}
This paper investigates the relationships among key elements of the scientific research network, namely articles, researchers, and journals. We introduce a novel approach to use semantic information through the HITS algorithm-based propagation of topic information in the network. The topic information is derived by using the Named Entity Recognition and Entity Linkage. In our case, MedCAT is used to extract the topics from the CORD19 Dataset, which is a corpus of
academic articles about COVID-19 and the coronavirus scientific network. Our approach focuses on the COVID-19 domain, utilizing the CORD-19 dataset to demonstrate the efficacy of integrating topic-related information within the citation framework. Through the application of a hybrid HITS algorithm, we show that incorporating topic data significantly influences article rankings, revealing deeper insights into the structure of the academic community. 
\end{abstract}

% keywords can be removed

\section{Introduction}
\label{intro}
The core elements of scientific research include articles, researchers, and institutions. Since scientific research is the cumulative effort of researchers to increase the understanding of the world around us, the relationships between these elements are as important as the scientific results themselves.

Gaining insight into the relationship between the core elements of scientific research can be useful for a variety of purposes, such as guiding scientific efforts toward better use of resources, inferring comparative results between fields of research, and better representing the importance of certain research fields and research groups.

Current scientific literature continues to grow at a rapid pace every day. Let alone being able to follow the growth of communities of which we are not a part, it has become very difficult to even find conferences, journals, or other prominent studies in our field. Naturally, examining the academic community in detail becomes a great burden for most young or experienced researchers, which results in missing out on promising researchers and useful works. To overcome this problem, most researchers represent the scientific literature as a wide network consisting of different entities such as researchers, papers, and journals with the help of some appropriate network measures \cite{Cai2018ScholarlyIA}, \cite{Walker2006RankingSP}, \cite{Wang2013RankingSA}, \cite{Zhang2019RankingSA}. In this paper, we aimed to analyze current literature and demonstrate different approaches to this problem with some practical applications. 

The academic writers, their studies, and the citation connection between them compose the scientific community, which forms a wide network of \textbf{authors} and \textbf{articles}. Authors are identified as the entities that create knowledge in the community through the articles they have published. The citation network which is derived from the published work is the most common representation of this knowledge, which is very simple yet effective in analyzing the communities. \textbf{Social Network Analysis}  is a way of measuring and mapping various aspects of relationships between different entities such as people, organizations, and groups \cite{Sweet2018SocialNA}. In the first step, we started our analysis from a simple representation of the network which is a graph of authors and articles. Then, we focus on the possible interpretations of the centrality metrics, PageRank, and its variations in real-world scenarios.

Apart from the approach above, it is clear that the proposed citation network lacks the semantic meaning of the published works \cite{Nart2015ACA}. Also, the representation of the topics is missing in the constructed graph of authors and their articles. Even though some solutions based on Natural Language Processing are available in the literature, most of these works require processing and analyzing the content of the published articles via their open-access files or abstracts \cite{Nart2015ACA}, \cite{Velardi2008ANC}. So, we propose a new and robust method to represent and retrieve the data in the scientific network by considering the topics as an entity in the research graph. The topics are derived from a pipeline based on Named Entity Recognition and Knowledge Base of the relevant field, which requires a focus on a specific domain to use the knowledge base effectively. Eventually, we showed the applicability of our method on a chosen domain and dataset, which are the COVID-19 and CORD19 Dataset.

Our motivation in this paper is to develop a way to provide the researchers with quantifiable information about the relationships between these elements so that they can be used for such purposes. This quantifiable information includes graph measures of individual elements of a graph as well as the graph measures of the whole graph.
Briefly, we propose a pipeline to create, analyze, and store the research network which consists of authors, articles, named entities, and relationships between them.

Moreover, we introduce the use of the MedCAT Concept Annotation Tool \cite{Kraljevic2020MultidomainCN}  and relevant medical ontology thanks to UMLS \cite{Lindberg1993TheUM}. Based on the heterogeneous network built with academic entities, we have conducted a bunch of experiments for the different link weighting schemes, on the other hand proposing a different approach for ground truth.
\section{Related Works}
\label{related}
The idea of Social Network Analysis was first proposed in 1969 by Philip Mayer and Julia C. Mitchell in their work on Urban situations in Central African Towns \cite{Mayer1970SocialNI}. Over time, Social Network Analysis has been used to understand relationships between different groups, organizations, communities, and any other possible network \cite{Su2019BibliometricSO}. Social Network analysis has an interdisciplinary nature as is in the intersection of sociology, mathematics, statistics, and computer science. In addition to these, the rise of Big Data in recent years has led to the analysis of communities in a more efficient way. For example, it has been a common approach to analyze connected social media data to detect misinformation or influential behaviors \cite{Ahmed2020COVID19AT}, \cite{Vassey2022EcigaretteBA}.

The scientific community is another domain that can be analyzed to derive meaningful information because of its interconnected nature.
One of the earliest works in the literature is the \textit{Impact Factor} proposed by Garfield (1972) in order to calculate the effectiveness and impact of the journals \cite{Garfield1972CitationAA}. He calculated the \textit{ImpactFactor} with the formula: $$\textit{ImpactFactor}(j,i) = A/B$$ where $A$ is the number of times articles published in journal j
in years $i-1$ and $i-2$ were cited in indexed journals and $B$ is the number of articles, reviews, proceedings or
notes published in journal j years $i-1$ and $i-2$. Following his works many researchers aimed to rank articles, authors, and journals based on their impact on the scientific community \cite{Cai2018ScholarlyIA}. 

Considering the article ranking methods, plenty of earlier works are derived from the PageRank algorithm \cite{Page1999ThePC}. Mostly this approach causes biased results due to the fact that the older papers, which have naturally higher citations than the newer ones, are assigned with higher ranking. CiteRank was proposed to remove the bias caused by PageRank by assigning more probability to new articles so that the random surfer model can choose these articles \cite{Walker2006RankingSP}. Additionally, FutureRank \cite{Sayyadi2009FutureRankRS} and P-Rank \cite{Yan2011PRankAI} algorithms were proposed in order to make use of different aspects such as time indicator, authorship, journal information, etc. P-Rank shifted the focus to understanding heterogeneous network representation of the entities \cite{Zhang2019RankingSA}. The heterogeneous network proposed in the P-Rank algorithm consists of author, article, and journal layers which propagate information among themselves to rank the specific article in the main network. 

Apart from these works, the HITS algorithm was proposed by Kleinberg. HITS algorithm uses authority and hub concepts to exploit the local structure of the network \cite{Kleinberg1999AuthoritativeSI}. The W-Rank then used both PageRank and HITS algorithms in order to utilize link weights based on citation and authorship relationships \cite{Zhang2019RankingSA}.

The existing solutions in the literature generally ignore the importance of the different edges in the heterogeneous network. Some researchers use topics for academic search by using topic modeling and its integration into the random walk framework \cite{Tang2008ATM}, however most of these methods lack of motivation to use topics in a weighting scheme to understand the nature of the community. Even though the time information is also used to evaluate link importance in citation relations in CiteRank algorithm \cite{Walker2006RankingSP}, it suffers from the absence of semantic meaning and heterogeneity. Also, we first present a way to assess the semantic meaning by approaching the topics as not only a similarity measure but also a part of the network.

\section{Methodology}
\label{methodology}
Given the nature of any research field, the network of scientific knowledge and researchers is immensely complex. Therefore, to make sense of how aspects of these graphs relate to each other, one would not only need the quantifying information on the graph but also how this information changes as the graph itself evolves.

There are some assumptions before building the network we proposed: 
\begin{itemize}
    \item Old articles tend to have higher citations, which leads to biased results in ranking algorithms.
    \item The articles in the prestigious journals tend to have be higher influence on the network without their momentary citation count.
    \item The prestigious authors tend to publish articles with a bigger influence.
    \item Important articles are cited by other important articles. The meaning of all the citations is not the same.
    \item The semantically similar papers tend to cite each other and such a citation has more importance if their topics are dominant in the network.
\end{itemize}

Considering these assumptions, our network structure is mainly based on the citation network which is composed of article nodes. The authors of and journal, if possible, of the article are connected to it with author and journal vertices. These attributes can be thought of as another layer of the network that is used to propagate information in prior works. The time information is also used to weight citation links between articles. Such network structure is very similar to the PageRank + HITS approaches we mentioned above with various weighting schemes. Differently, we add the topic layer, similar to the author and journal layers, to the network by also exploring the influence of topics on citation weights. Eventually, a lightweight semantic network can be a part of the graph, which helps to analyze semantic relationships between articles in the network. We also believe that such an approach enables us to exploit topic-based search and understand the topic-wise prestige of articles in the network. 

Before starting to theoretical groundings of the algorithm, we are stating that the works \cite{Wang2013RankingSA} and \cite{Zhang2019RankingSA} are used as a basis for our PageRank + HITS algorithm. Our main contribution is introducing the concept of the \textit{topic} for such an algorithm.

\subsection{Heterogeneous Network}
The heterogeneous approach has become a common approach when investigating the scientific communities. From the formal perspective a heterogeneous graph can be defined as:
\begin{multline}
$$
G(V,E)=(V_{ar} \cup V_{au} \cup V_{ju},\\ E_{ar-ar} \cup E_{ar-au} \cup E_{ar-ju}) 
$$ 
\end{multline}
where $V_{ar}$, $V_{au}$ and $V_{ju}$ are the vertices of article, author and journal networks, whereas $E_{ar-ar}$, $E_{ar-au}$ and $E_{ar-ju}$ are the edges respectively. Based on this definition, we consider the topics as a part of the graph by adding vertices between articles and topics. Unlike the author and journal layer, the topics have a connection among themselves, which is analyzed later. So eventually, our heterogeneous network has the formula:

\begin{multline}
$$
G(V,E)=(V_{ar} \cup V_{au} \cup V_{ju} \cup V_{tp} , \\
E_{ar-ar} \cup E_{ar-au} \cup E_{ar-ju}  \cup E_{ar-tp}   \cup E_{tp-tp} ) 
$$
\end{multline}
where $V_{tp}$ represents the vertices of the topic network, whereas $E_{ar-tp}$ stands for the edges from articles to topics. $E_{tp-tp}$ is a term added for the topic network which has a hierarchy derived from the thesaurus and ontology of biomedical concepts tanks to Unified Medical Language System (UMLS).

\begin{figure}
    \centering
    \includegraphics[width=1\linewidth]{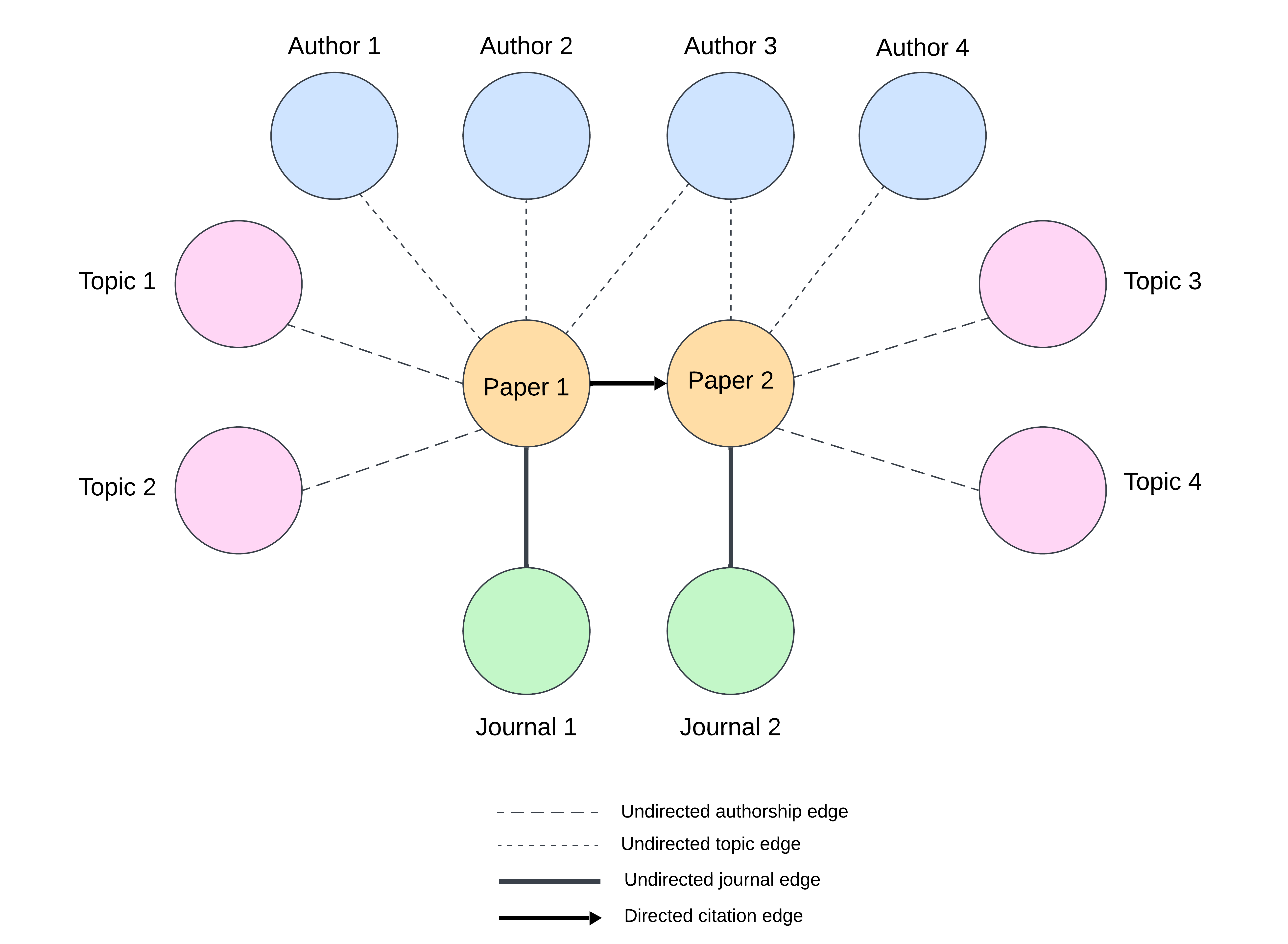}
    \caption{A schematic of the heterogeneous network}
    \label{fig:enter-label}
\end{figure}
\subsection{Link Weighting}
Depending on the type of vertices, different weighting schemes can be employed. At this step, each relationship type is analyzed separately.
\subsubsection{Article-Author}
It is quite obvious that authors contribute differently to their published works. Although in some disciplines such as computer science, the ordering of the authors implies the importance and contribution, it is not a standardized approach in the scientific community. Additionally, it can lead to underestimating the contribution of the authors, which is a serious issue in the academic community. Some research includes $H-Index$ based solutions however this approach can assign lower ranks to young researchers who have lower $H-Index$. 

\subsubsection{Article-Journal}
The journals tend to publish similar quality articles in line with their own prestige, so we believe that we reach the articles published with equal probability starting from a specific journal. There are only two possibilities in the article-journal networks, which are \textit{publish} or \textit{not publish} \cite{Zhang2019RankingSA}. 
\subsubsection{Article-Article}
The citation network has more information than other sub-networks, which stems from the complex and versatile nature of the citation relationship. In addition to the graph-based approach, many works explore Natural Language Processing techniques to understand the importance of the citation. However, most of these works require a huge amount of effort and data to train and classify the relevant models to understand whether a citation is influential or not. We use a hybrid approach by combining graph attributes and processing the abstract information to find similarities between articles. Using the abstract is more robust, efficient, and fast than trying to find citation text in the document. Also, it prevents us from being limited to articles that are only open-access PDFs.
Briefly, the citation weight should have two different parts which are semantic-based similarity and network-based similarity \cite{Zhang2019RankingSA}. Based on the work proposed by Zhang et al. we improve the network-based similarity by adding parameters related to authorship and journals in which published.

\begin{equation}
\begin{split}
    S_n(P_1,P_2) = \alpha \cdot \frac{|(In_{P_1} \bigcup Out_{P_1})\bigcap (In_{P_2} \bigcup Out_{P_2})|}{\sqrt{|In_{P_1} \bigcup Out_{P_1}|\times |In_{P_2} \bigcup Out_{P_2}|}} \\
    + \beta \cdot \frac{|A_{P_1}\bigcup A_{P_2}|}{\sqrt{|A_{P_1}|\times |A_{P_2}|}} \\
    + \gamma \cdot J_{P1-P2} 
\end{split}
\end{equation}

where $In_P$ and $Out_P$ are incoming and outgoing links, whereas $A_P$ is the authors of the the article $P$. $J_{P1-P2}$ can be $1$ or $0$ depending on whether articles $P_1$ and $P_2$ were published in the same journal or not. The coefficients $\alpha$, $\beta$ and $\gamma$ are $0.6$, $0.3$ and $0.1$, respectively.

We analyzed the topic related weighting in next chapter.

\subsection{Topic Linking and Semantic Weighting}
Having a graph-structured representation of the research world allows the addition of explicit connections to other graph-structured knowledge representations. One prime example of such knowledge representations is ontologies. The possible connections between a selected ontology and, the Unified Medical Language System (UMLS), have been investigated. The connections between the papers and the UMLS concepts are constructed by passing the abstracts of the papers through a named entity recognizer called MedCAT.

In this paper, we propose a method that is based on Named Entity Recognition and Linking (NER+L) to extract the relevant concepts from the article abstracts based on MedCAT and the Unified Medical Language System.

MedCAT \cite{Kraljevic2020MultidomainCN} is a content annotation tool based on Word2Vec embeddings that can be used to extract information from medical documents to link them to medical ontologies such as UMLS \cite{Lindberg1993TheUM}. 

Here our semantic weighting formula:
\begin{equation}
\begin{split}
    S_s(P_1, P_2)= \frac{T_{P_1} \cap T_{P_2}}{\sqrt{|T_{P_1}| \cdot |T_{P_2}|}}
\end{split}
\end{equation} where $T_{P_1}$ is the topics of the article i.

Resulting final weight is:

\begin{equation}
    \begin{split}
        S(P_1, P_2) = \alpha \cdot S_n + \beta \cdot S_s
    \end{split}
\end{equation} where:\\
$\alpha = e^{\lambda (S_n(P_1,P_2)-\tau_1)}$ and
$\beta = e^{\lambda (S_s(P_1,P_2)-\tau_2)}$
where $\lambda=6$ and $\tau_1$ and $\tau_2$ are median values of $S_n$ and $S_s$.

\subsection{Ranking Algorithm}
Based on the HITS algorithm \cite{Kleinberg1999AuthoritativeSI}, we use a weighted iteration and updates of authorities and hubs in the network.
The previous research \cite{Wang2013RankingSA} and \cite{Zhang2019RankingSA} are followed when defining the scores in the algorithm. Naturally, due to the newly introduced topic network, the score functions are slightly different.
\subsubsection{Hub Scores}

The hub scores in the scientific network can be interpreted as the quality and impact of the hubs to which the relevant entity belongs. We analyzed the hub scores of articles, authors, journals, and topics in this direction. We followed the slight derivation of the HITS algorithm by normalizing the hub score with the number of the links \cite{Wang2013RankingSA}, because of the risk that authors and journals that publish huge numbers of articles can dominate the hubs.
In addition to the work of Wang et al., we present the hub scores of the topics by considering the topics as a part of the heterogeneous network. Apart from these, to understand the current state of the network we also followed the time-aware approaches similar to prior works \cite{Wang2013RankingSA}. Time-aware weights for article-author, article-journal, and article-article links are needed to score the hubs.

The hub score of an author $i$:
\begin{equation}
\begin{gathered}
    $$
    H(A_i) = \frac{\sum_{P_j \in L_i} w_{t}(i) \cdot A(P_j)}{|L_i|}
    $$
\end{gathered}
\end{equation}

where $L_i$ is the articles published by author $i$, $A(P_j)$ is the authority score of article j, $w_{t}(i)$ is the time-aware weight between author $i$ and article $j$, and $H(A_i)$ is the hub score of the author $i$. Then all the hub scores of authors are normalized to $1$.

The hub score of a journal $i$:
\begin{equation}
\begin{gathered}
    $$
    H(J_i) = \frac{\sum_{P_j \in K_i} w_{t}(i) \cdot A(P_j)}{|K_i|}
    $$
\end{gathered}
\end{equation}

where $K_i$ is the articles published in journal $i$, $A(P_j)$ is the authority score of article j, $w_{t}(i)$ is the time-aware weight between journal $i$ and article $j$, and $H(J_i)$ is the hub score of the journal $i$. Then all the hub scores of journals are normalized to $1$.

The hub score of a topic $i$:
\begin{equation}
\begin{gathered}
    $$
    H(T_i) = \frac{\sum_{P_j \in M_i} w_{t}(i) \cdot A(P_j)}{|M_i|}
    $$
\end{gathered}
\end{equation}

where $M_i$ is the articles published related to topic $i$, $A(P_j)$ is the authority score of article j, $w_{t}(i)$ is the weight between topic $i$ and article $j$, and $H(T_i)$ is the hub score of the topic $i$. Then all the hub scores of topics are normalized to $1$.

The hub score of an article $i$:
\begin{equation}
\begin{gathered}
    $$
    H(P_i) = \frac{\sum_{P_j \in N_i} w_{t}(j) \cdot A(P_j)}{|N_i|}
    $$
\end{gathered}
\end{equation}

where $N_i$ is the articles cites or cited by the article $i$, $A(P_j)$ is the authority score of article j, $w_{t}(i)$ is the weight between article $i$ and article $j$, and $H(P_i)$ is the hub score of the article $i$. Then all the hub scores of articles are normalized to $1$.

The time aware weighting for the hubs:
\begin{equation}
    w_t(i) = a^{T_{current} - T_i}
\end{equation} where a = 2.

\subsubsection{Authority Scores}
The authority scores of the papers can be calculated as:
\begin{equation}
\begin{split}
    AS(P_i) = \alpha \cdot PageRank(P_i)\\
             + \beta \cdot Author(P_i) \\
             + \gamma \cdot Journal(P_i) \\
             + \delta \cdot Topic(P_i) \\
             + \Omega \cdot Article(P_i) \\
             + \Sigma  \cdot P^{Time}_i \\
             + (1-\alpha-\beta-\gamma-\delta-\omega-\sigma) \cdot \frac{1}{N_p}
\end{split}
\end{equation}
where $\beta, \gamma, \delta, \omega,$ and $\sigma$ are constant parameters for the authority scores for the Author, Journal, Citation authority scores. 

These authority scores are calculated by following formulas:

\textbf{PageRank Score} 
\begin{equation}
    PageRank(P_i) = \sum_{P_j\in In(P_i)} \frac{w_{i,j}}{\sum_{P_m \in Out(P_j)}w_{j,m}}AS(P_j)
\end{equation} where $w_{i,j}$ is the weight between articles.

\textbf{Author Authority Score}
    \begin{equation}
        Author(P_i) = \frac{1}{Z(A)} \sum_{A_j\in Neighbours(P_i)} w_{t2}(i) H(A_j)
    \end{equation} where $Neighbours(P_i)$ is the authors of the paper $P_i$, $H(A_j)$ is the hub score of the author j, and $Z(A)$ is normalized of sum of scores transferred from all authors to all papers.

\textbf{Journal Authority Score}
    \begin{equation}
        Journal(P_i) = \frac{1}{Z(J_i)} \sum_{J_j\in Neighbours(P_i)} w_{t2}(i) H(J_j)
    \end{equation} where $Neighbours(P_i)$ is the journals of the paper $P_i$, $H(J_j)$ is the hub score of the journal j, and $Z(J)$ is normalized of sum of scores transferred from all journals to all papers.

\textbf{Topic Authority Score}
    \begin{equation}
        Topic(P_i) = \frac{1}{Z(T)} \sum_{T_j\in Neighbours(P_i)} w_{t2}(i) H(J_j)
    \end{equation} where $Neighbours(P_i)$ is the topics of the paper $P_i$, $H(T_j)$ is the hub score of the topic j, and $Z(J)$ is normalized of sum of scores transferred from all topics to all papers.

\textbf{Article Authority Score}
    \begin{equation}
        Article(P_i) = \frac{1}{Z(P)} \sum_{P_j\in Neighbours(P_i)} w_{t2}(i) H(P_j)
    \end{equation} where $Neighbours(P_i)$ is the neighbours (citing and cited by) of the paper $P_i$, $H(P_j)$ is the hub score of the article j, and $Z(J)$ is normalized of sum of scores transferred from all topics to all papers.

\textbf{Time Value}
\begin{equation}
     P^{Time}_i = e^{-p* (T_{current} - T_i)}
\end{equation}

The time aware weighting for the authority scores:
\begin{equation}
    w_{t2}(i) = \frac{1}{1 + b \cdot (T_{current} - T_i)}
\end{equation} where b = 1.

The algorithm in the previous work \cite{Zhang2019RankingSA} is employed to calculate the authority scores of articles with the score functions above.
\section{Experiments and Results}
\label{experiment}
\subsection{Dataset}
We use the public dataset CORD19 which is a corpus of academic articles about COVID-19 and coronavirus scientific network published by Semantic Scholar Team at the Allen Institute for AI \cite{wang-etal-2020-cord}. The final version was released on June 2, 2022 with 1M+ papers and around 370k full text support. We have performed a pre-processing that eliminates the duplicate articles, articles with no year information or journal information and broken rows. Also the dataset has not any information about the citations among the articles. To create a graph representation of CORD19 Dataset, we have used Semantic Scholar \cite{Kinney2023TheSS} to fetch the citation information related to each articles. By doing this operation, we also aimed to synchronize dataset with the Semantic Scholar Academic Graph API. Eventually, we have created and published a graph dataset containing 728675 articles, 2210182 authors, and 5875663 citations. The dataset represents the citation network of the CORD19 Dataset on January 2024.

For experiments, it would be computationally challenging to use all the graphs to find the the related metrics, so we have performed our analysis on a subset of the graph containing 19981 articles, 121431 authors, 2925 journals, and 209788 citations, by considering the papers with citation 20 or more. After paper selection, we investigate the citations among these new paper sets. This process can be though of as a cutting of papers with low citations. The resulting graph is not so dense because only links between them are considered. 

The semantic network was created using the MedCAT \cite{Kraljevic2020MultidomainCN} outputs of the article abstracts. Here we follow the Unified Medical Language System \cite{Lindberg1993TheUM} and the unique identifiers in the Metathesaurus Corpus. The extracted \textbf{Concept Unique Identifiers (CUI)} are used to build the semantic network by creating the links to related articles by setting an accuracy threshold, \%75 in our pipeline. Some Concept Unique Identifiers are discarded by considering their type ids, see Appendix \ref{appendix:main_terms} and \ref{appendix:discarded_terms}.
\subsubsection{Ground Truth and Evaluation Criteria}
Finding a ground truth to analyze the academic networks is a common issue that most researchers tried to answer \cite{Su2019BibliometricSO}. In our paper, we do not follow the prior ground truth because it would be contradictory to use metrics and algorithms based on old approaches as ground truth while trying to find a novel metric or trying to show that new parameters change the results. Since there was no human-based annotation on the dataset we worked with, we made a comparative evaluation criterion. In this case, we tried to analyze how different parameters in the algorithm change the result and what different semantic network causes.

Briefly, we investigated the change of most ranked papers and correlation coefficient by changing the parameters in order to see their effect on the main algorithm. The distance and irrelevance of the settings to the basis PageRank dominated algorithm can been as a differentiation from the traditional PageRank solutions.

We assess the correlation through Spearman's rank correlation \cite{Myers1991ResearchDA}:
\[
\rho = \frac{\sum_i \left( R_1(P_i) - \overline{R}_1 \right) \left( R_2(P_i) - \overline{R}_2 \right)}
{\sqrt{\sum_i \left( R_1(P_i) - \overline{R}_1 \right)^2 \sum_i \left( R_2(P_i) - \overline{R}_2 \right)^2}}
\]
where $R(P_i)$ are position of a specific in each lists, $\overline{R}$ are average ranks of papers in these lists. When there is a tie, the rank position is the average rank of all the ties.

\begin{table*}[t]
\centering
\tiny

\caption{The top 10 ranked articles }

\begin{tabular}{|p{0.19\textwidth}|p{0.19\textwidth}|p{0.19\textwidth}|p{0.19\textwidth}|p{0.19\textwidth}|}
\hline
\textbf{Topic} & \textbf{Author} & \textbf{Journal} & \textbf{Article} & \textbf{PageRank} \\
\hline
The clinical pathology of severe acute respiratory syndrome (SARS): a report from China & Isolation from Man of "Avian Infectious Bronchitis Virus-like" Viruses (Coronaviruses*) similar to 229E Virus, with Some Epidemiological Observations & Clinical Characteristics of Coronavirus Disease 2019 in China & Early Transmission Dynamics in Wuhan, China, of Novel Coronavirus–Infected Pneumonia & Identification of a novel coronavirus in patients with severe acute respiratory syndrome. \\
\hline
A novel coronavirus associated with severe acute respiratory syndrome. & Antigenic relationships among the coronaviruses of man and between human and animal coronaviruses. & Ultrastructural analysis of SARS-CoV-2 interactions with the host cell via high resolution scanning electron microscopy & A Novel Coronavirus from Patients with Pneumonia in China, 2019 & Identification of severe acute respiratory syndrome in Canada. \\
\hline
Identification of severe acute respiratory syndrome in Canada. & SEROEPIDEMIOLOGIC STUDIES OF CORONAVIRUS INFECTION IN ADULTS AND CHILDREN1 & Stability of SARS-CoV-2 on critical personal protective equipment & The Human Respiratory System and its Microbiome at a Glimpse & Evidence of human metapneumovirus in Australian children \\
\hline
A cluster of cases of severe acute respiratory syndrome in Hong Kong. & Clinical Characteristics of Coronavirus Disease 2019 in China & Genomic mutations and changes in protein secondary structure and solvent accessibility of SARS-CoV-2 (COVID-19 virus) & Antigenic relationships among the coronaviruses of man and between human and animal coronaviruses. & A Novel Coronavirus from Patients with Pneumonia in China, 2019 \\
\hline
Clinical Characteristics of Coronavirus Disease 2019 in China & A Novel Coronavirus from Patients with Pneumonia in China, 2019 & Forecasting the spread of COVID-19 under different reopening strategies & First Case of 2019 Novel Coronavirus in the United States & Importation and Human-to-Human Transmission of a Novel Coronavirus in Vietnam \\
\hline
A Novel Coronavirus from Patients with Pneumonia in China, 2019 & Visualization by Immune Electron Microscopy of a 27-nm Particle Associated with Acute Infectious Nonbacterial Gastroenteritis & A Novel Coronavirus from Patients with Pneumonia in China, 2019 & The Impact of COVID-19 on Italy: A Lesson for the Future & Assessing spread risk of Wuhan novel coronavirus within and beyond China, January-April 2020: a travel network-based modelling study \\
\hline
Evidence of human metapneumovirus in Australian children & Middle East Respiratory Syndrome Coronavirus (MERS-CoV): A Perpetual Challenge & Early Transmission Dynamics in Wuhan, China, of Novel Coronavirus–Infected Pneumonia & Prophylactic and therapeutic remdesivir (GS-5734) treatment in the rhesus macaque model of MERS-CoV infection & First Case of 2019 Novel Coronavirus in the United States \\
\hline
Early Transmission Dynamics in Wuhan, China, of Novel Coronavirus–Infected Pneumonia & Studies With Human Coronaviruses II. Some Properties of Strains 229E and OC43 & Characteristics of and Important Lessons From the Coronavirus Disease 2019 (COVID-19) Outbreak in China: Summary of a Report of 72 314 Cases From the Chinese Center for Disease Control and Prevention. & SEROEPIDEMIOLOGIC STUDIES OF CORONAVIRUS INFECTION IN ADULTS AND CHILDREN1 & A novel coronavirus associated with severe acute respiratory syndrome. \\
\hline
First Case of 2019 Novel Coronavirus in the United States & Early Transmission Dynamics in Wuhan, China, of Novel Coronavirus–Infected Pneumonia & [Asymptomatic infection of COVID-19 and its challenge to epidemic prevention and control]. & Incubation period of 2019 novel coronavirus (2019-nCoV) infections among travellers from Wuhan, China, 20–28 January 2020 & Early Transmission Dynamics in Wuhan, China, of Novel Coronavirus–Infected Pneumonia \\
\hline
Characteristics of and Important Lessons From the Coronavirus Disease 2019 (COVID-19) Outbreak in China: Summary of a Report of 72 314 Cases From the Chinese Center for Disease Control and Prevention. & First Case of 2019 Novel Coronavirus in the United States & First Case of 2019 Novel Coronavirus in the United States & Human Coronavirus in Hospitalized Children With Respiratory Tract Infections: A 9-Year Population-Based Study From Norway & A major outbreak of severe acute respiratory syndrome in Hong Kong. \\
\hline
\end{tabular}
\label{tab:covid-papers}
\end{table*}
\subsubsection{Experiment Setup}
We conducted experiments by changing the parameters in the algorithm to see their behaviours.  6 different kind of metrics:
\begin{itemize}
    \item Alpha: traditional PageRank scores of the articles in citation network
    \item Beta: Authority scores propagated from author-article network
    \item Gamma: Authority scores propagated from journal-article network
    \item Delta: Authority scores propagated from topic-article network
    \item Omega: Authority scores propagated from article-article network, namely citation network
    \item Sigma: Time Factor of the articles
\end{itemize}

Among our experiments, we have chosen 12 settings to show the relationship between semantic network and the other networks. We ensured that equation $\alpha+\beta+\gamma+\delta+\omega+\sigma=1$ is met in all settings with the $0.1$ random jump probability coming from $1-(\alpha+\beta+\gamma+\delta+\omega+\sigma)$. See the appendix to see exact values of parameters for the 12 different settings.

Apart from these, based on the prior works mentioned above, we set the time factor to $0.1$ in all the calculations.

We have created various configurations through changing the coefficients above. 
\begin{itemize}
    \item \textbf{PageRank}: the PageRank score of the articles in the network. Base metric for the following combinations.
    \item \textbf{Author Information}: Author-article network information is added to network. It represent the authority scores coming from authors to articles.
    \item \textbf{Topic Information}: Topic-article network information is added to network. It represent the authority scores coming from topics to articles. 
    \item \textbf{Journal Information}:  Journal-article network information is added to network. It represent the authority scores coming from journals to articles.
    \item \textbf{Article Information}:  Citation network information is added to network. It represent the authority scores coming from articles to articles.
\end{itemize}

You can find the code and dataset used in experiments in the Github\footnote{\quad \url{https://github.com/mehmetemreakbulut/content-aware-analysis-of-scholarly-networks}}\space \space .
\newpage
\subsection{Experiments and Result}

\begin{figure}[h]
\includegraphics[width=8cm]{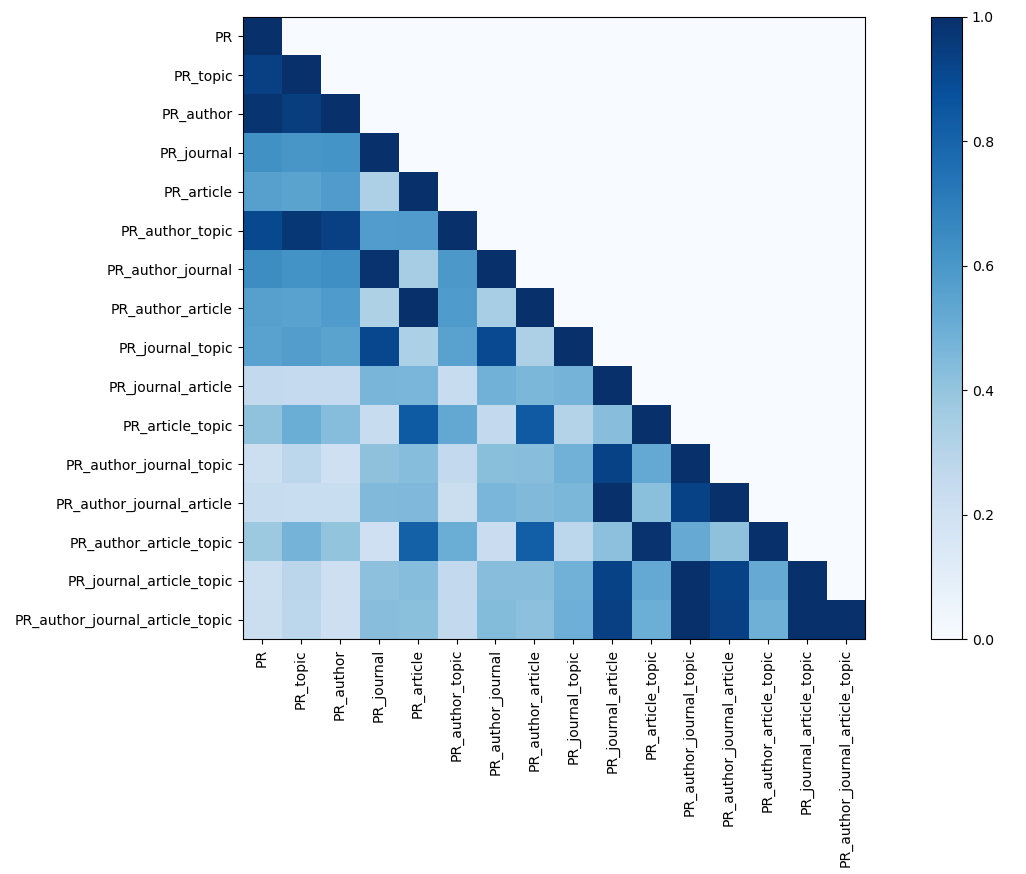}
\caption{Correlation scores of different settings}
\label{fig:results}
\end{figure}

The traditional PageRank algorithm is used as a baseline for the experiments. Even though it is not a good choice to be a ground truth, understanding which networks have lower correlation may help us to differentiate the behaviors of the article, author, journal, and topic networks.

In Figure 1, the correlation scores decrease as the new metric is added to the calculation function. However, adding only a topic or author to the calculation does not change results as much as adding journal or article information. Nevertheless, topic-weighted calculation causes different results for the versions with no topic information considering their similarity to the baseline PageRank-only calculation. For instance, $PR\_journal$ setting has $0.62$ correlation score with PageRank, which is very low considering that it still has PageRank coefficient $\alpha=0.5$. Adding the topic value leads to a $0.55$ correlation score. Similarly, $PR\_article$ has $0.56$ correlation, while $PR\_article\_topic$ is much lower with the correlation $0.41$. Also, consider the settings $PR\_author\_journal$ with $0.64$ correlation to PageRank and $PR\_journal\_topic$ with $0.55$ correlation. At this step, it is clear that the topic network propagates more information than the author network considering its effect on the basic settings. The article information is the strongest among the settings. Naturally, it has a very low correlation with the baseline setting because of the highly informative nature of the citation network. On the other hand, we believe that the sparsity of the journal network is the main reason for the effect of journal information.

Apart from the basic settings, the complex settings in which at least 3 networks are used also produce similar results. We investigated the correlation of the settings concerning the setting in which every network is used, namely $PR\_author\_journal\_article\_topic$. Its lowest correlation among the complex settings is with the $PR\_author\_article\_topic$, which is similar to the above results indicating the importance of journal information. Even though other correlations are similar, the second highest drop in correlation occurs when we discard the topic information from this setting.

\subsubsection{Top Articles}
\begin{table}[htbp]
\centering
\begin{tabular}{|l|c|c|c|c|c|}
\hline
\textbf{Setting} & \textbf{Topic} & \textbf{Author} & \textbf{Journal} & \textbf{Article} \\

\hline
\textbf{Author} & 84 & -- & -- & -- \\
\hline
\textbf{Journal} & 77 & 71 & -- & -- \\
\hline
\textbf{Article} & 16 & 18 & 15 & -- \\
\hline
\textbf{PR} & 55 & 58 & 42 & 14 \\
\hline
\end{tabular}
\vspace{2pt}
\caption{Common papers among different settings}
\label{tab:common-papers}
\end{table}

In Table \ref{tab:common-papers}, we have analyzed the top 100 results with respect to different settings. At this time, we enable a network to dominate the results by assigning it to the maximum possible coefficient value and nullifying the others.

Similar to the previous experiment, the article network hasn't much common paper with PageRank dominated setting. Moreover, the article network has so little common paper compared to all the other settings. Also, the journal-dominated setting has more common articles with the topic-dominated setting than the author-dominated setting. This behavior demonstrates the journal-topic relationship is stronger than the journal-author relationship, which also can be derived from the correlation scores in previous experiment. Using topic instead of author together with journal information causes more differentiated results than the PageRank-dominated settings.

Table \ref{tab:covid-papers} shows the top 10 ranked articles from 5 different settings. In the topic-dominated setting column, it is easy to spot that most of the articles have broader topics than the best-ranked articles of other settings. They mainly include \textbf{Severe Acute Respiratory Syndrome (SARS)} text in their title, naturally in their abstracts. Also, these articles mostly tried to understand and explain the literature. As we observed, general terms such as \textit{"respiratory"}, \textit{"acute"} and \textit{"cell"} tend to occur frequently in the abstract of these articles. Based on this phenomenon, we can conclude that the topic-dominated settings tend to give results favoring the literature reviews, reports, and broader articles. 

In the table, there are some papers among the top papers with a very low number of citations, such as \textit{"Ultrastructural analysis of SARS-CoV-2 interactions with the host cell via high resolution scanning electron microscopy1}. Along with the papers in the rank $3$ and $4$, three of the top 5 articles in journal dominated setting belong to journal \textit{Scientific Reports, Nature}. Considering the academic prestige of Nature, it is not a surprise to see these papers when the journal information dominates the information flow in the network.

\subsubsection{Effect of Citation}
See the Appendix \ref{appendix:citation}

The effect of a number of citations is also important to understand the different behavior of the settings. To achieve this, we calculate Spearman's rank correlation coefficient between the settings and the citation count-based ranking.

\begin{table}[htbp]
\centering
\begin{tabular}{|l|c|}
\hline
\textbf{Metric} & \textbf{Correlation} \\
\hline
Topic vs Citation Count & 0.2525 \\
\hline
Author vs Citation Count & 0.2735 \\
\hline
Journal vs Citation Count & 0.2789 \\
\hline
Article vs Citation Count & 0.2575 \\
\hline
PR vs Citation Count & 0.6151 \\
\hline
\end{tabular}
\vspace{2pt}
\caption{Correlation of settings with citation count}
\end{table}

It is a widely known fact that PageRank-based solutions can be biased because highly cited papers can dominate the scores. The purpose of this research is to propose a framework to investigate the network from different perspectives without biased citation-based approaches. From this point of view, the correlation between topic and citation count is lower than the author, journal, and article.

\section{Conclusion and Future Work}
In this paper, we have explored the possible ways of using the topic-related information by propagating it with a hybrid HITS algorithm following the prior works. We have shown that proper parameter selection, with respect to author, journal, topic, or article network, can lead to different rankings for articles. At this point, we see that the topic network has more power to change the article rankings than the author network. Also, adding topic information leads to a lower correlation to the number of citations compared to other networks. This phenomenon arises from not only propagation from the topic network but also semantic weighting of the citation links.

We believe that using a topic network that also has interconnected links between their nodes such as subclass relation, can lead to better results with the help of a strong representation of the semantic relationships. In this case, using the UMLS Metathesaurus \cite{Aronson2001EffectiveMO} knowledge graph can be a suitable future work beyond our research.

A significant contribution would be the proposing of a topic matrix in the calculation to find the most ranked papers for a given topic at any time, which can improve the quality of the semantic ranking in the network.
\label{conclusion}

% keywords can be removed


\bibliography{references}
\clearpage

\begin{appendices}
\section{Number of citation of Top Papers}
\label{appendix:citation}
\begin{supertabular}{@{} p{15cm} c @{}} 
    \toprule
    \textbf{Title} & \textbf{Citations} \\ 
    \midrule

Identification of a novel coronavirus in patients with severe acute respiratory syndrome. & 759 \\ 
Genomic mutations and changes in protein secondary structure and solvent accessibility of SARS-CoV-2 (COVID-19 virus) & 3 \\ 
Assessing spread risk of Wuhan novel coronavirus within and beyond China, January-April 2020: a travel network-based modelling study & 21 \\  
Studies With Human Coronaviruses II. Some Properties of Strains 229E and OC43 & 5 \\ 
A novel coronavirus associated with severe acute respiratory syndrome. & 727 \\ 
Early Transmission Dynamics in Wuhan, China, of Novel Coronavirus–Infected Pneumonia & 1104 \\ 
The clinical pathology of severe acute respiratory syndrome (SARS): a report from China & 125 \\ 
Characteristics of and Important Lessons From the Coronavirus Disease 2019 (COVID-19) Outbreak in China: Summary of a Report of 72,314 Cases From the Chinese Center for Disease Control and Prevention. & 1043 \\ 
Antigenic relationships among the coronaviruses of man and between human and animal coronaviruses. & 24 \\ 
Prophylactic and therapeutic remdesivir (GS-5734) treatment in the rhesus macaque model of MERS-CoV infection & 85 \\ 
Asymptomatic infection of COVID-19 and its challenge to epidemic prevention and control. & 2481 \\ 
Incubation period of 2019 novel coronavirus (2019-nCoV) infections among travellers from Wuhan, China, 20–28 January 2020 & 125 \\ 
A cluster of cases of severe acute respiratory syndrome in Hong Kong. & 195 \\ 
Ultrastructural analysis of SARS-CoV-2 interactions with the host cell via high resolution scanning electron microscopy & 2 \\ 
The Impact of COVID-19 on Italy: A Lesson for the Future & 5 \\ 
Visualization by Immune Electron Microscopy of a 27-nm Particle Associated with Acute Infectious Nonbacterial Gastroenteritis & 3 \\ 
Importation and Human-to-Human Transmission of a Novel Coronavirus in Vietnam & 130 \\ 
Isolation from Man of “Avian Infectious Bronchitis Virus-like” Viruses (Coronaviruses*) similar to 229E Virus, with Some Epidemiological Observations & 18 \\ 
The Human Respiratory System and its Microbiome at a Glimpse & 1 \\ 
A major outbreak of severe acute respiratory syndrome in Hong Kong. & 408 \\ 
Human Coronavirus in Hospitalized Children With Respiratory Tract Infections: A 9-Year Population-Based Study From Norway & 10 \\ 
SEROEPIDEMIOLOGIC STUDIES OF CORONAVIRUS INFECTION IN ADULTS AND CHILDREN1 & 45 \\ 
A Novel Coronavirus from Patients with Pneumonia in China, 2019 & 1563 \\ 
Evidence of human metapneumovirus in Australian children & 17 \\ 
Clinical Characteristics of Coronavirus Disease 2019 in China & 1575 \\ 
Forecasting the spread of COVID-19 under different reopening strategies & 0 \\ 
Identification of severe acute respiratory syndrome in Canada. & 232 \\ 
Middle East Respiratory Syndrome Coronavirus (MERS-CoV): A Perpetual Challenge & 7 \\ 
Stability of SARS-CoV-2 on critical personal protective equipment & 2 \\ 
First Case of 2019 Novel Coronavirus in the United States & 524 \\ 

    \bottomrule
\end{supertabular}

\clearpage
\section{Parameters and Settings}
\begin{center}
\begin{supertabular}{@{} lcccccc @{}}
    \toprule
    \textbf{Metric}                       & \textbf{$\alpha$} & \textbf{$\beta$} & \textbf{$\gamma$} & \textbf{$\delta$} & \textbf{$\omega$} & \textbf{$\sigma$} \\ 
    \midrule

PR                                   & 0.8        & 0.0        & 0.0        & 0.0        & 0.0        & 0.1        \\ 
PR\_author                            & 0.5        & 0.3        & 0.0        & 0.0        & 0.0        & 0.1        \\ 
PR\_journal                           & 0.5        & 0.0        & 0.3        & 0.0        & 0.0        & 0.1        \\ 
PR\_article                           & 0.5        & 0.0        & 0.0        & 0.0        & 0.3        & 0.1        \\ 
PR\_author\_journal                  & 0.4        & 0.2        & 0.2        & 0.0        & 0.0        & 0.1        \\ 
PR\_author\_article                  & 0.4        & 0.2        & 0.0        & 0.0        & 0.2        & 0.1        \\ 
PR\_journal\_article                 & 0.4        & 0.0        & 0.2        & 0.0        & 0.2        & 0.1        \\ 
PR\_author\_journal\_article         & 0.2        & 0.2        & 0.2        & 0.0        & 0.2        & 0.1        \\ 
PR\_topic                             & 0.5        & 0.0        & 0.0        & 0.3        & 0.0        & 0.1        \\ 
PR\_author\_topic                    & 0.3        & 0.2        & 0.0        & 0.3        & 0.0        & 0.1        \\ 
PR\_journal\_topic                   & 0.3        & 0.0        & 0.2        & 0.3        & 0.0        & 0.1        \\ 
PR\_article\_topic                   & 0.3        & 0.0        & 0.0        & 0.3        & 0.2        & 0.1        \\ 
PR\_author\_journal\_topic           & 0.2        & 0.1        & 0.1        & 0.3        & 0.2        & 0.1        \\ 
PR\_author\_article\_topic           & 0.2        & 0.1        & 0.0        & 0.3        & 0.2        & 0.1        \\ 
PR\_journal\_article\_topic          & 0.2        & 0.0        & 0.1        & 0.3        & 0.2        & 0.1        \\ 
PR\_author\_journal\_article\_topic   & 0.2        & 0.133      & 0.133      & 0.2        & 0.133      & 0.1        \\ 
Topic Dominated                      & 0.1        & 0.0        & 0.0        & 0.7        & 0.0        & 0.1        \\ 
Author Dominated                     & 0.1        & 0.7        & 0.0        & 0.0        & 0.0        & 0.1        \\ 
Journal Dominated                    & 0.1        & 0.0        & 0.7        & 0.0        & 0.0        & 0.1        \\ 
Article Dominated                    & 0.1        & 0.0        & 0.0        & 0.7        & 0.0        & 0.1        \\ 
  \bottomrule
\end{supertabular}
\end{center}
\clearpage

\clearpage
\section{Main Terms Table}
\label{appendix:main_terms}
\begin{supertabular}{@{} p{10cm} c c @{}}
    \toprule
    \textbf{Description} & \textbf{Code} & \textbf{ID} \\ 
    \midrule
    Amino Acid, Peptide, or Protein          & aapp  & T116 \\ 
    Acquired Abnormality                     & acab  & T020 \\ 
    Age Group                                & aggp  & T100 \\ 
    Amino Acid Sequence                      & amas  & T087 \\ 
    Amphibian                                & amph  & T011 \\ 
    Anatomical Abnormality                   & anab  & T190 \\ 
    Animal                                   & anim  & T008 \\ 
    Anatomical Structure                     & anst  & T017 \\ 
    Antibiotic                               & antb  & T195 \\ 
    Archaeon                                 & arch  & T194 \\ 
    Biologically Active Substance            & bacs  & T123 \\ 
    Bacterium                                & bact  & T007 \\ 
    Body Substance                           & bdsu  & T031 \\ 
    Body System                              & bdsy  & T022 \\ 
    Biologic Function                        & biof  & T038 \\ 
    Bird                                     & bird  & T012 \\ 
    Body Location or Region                  & blor  & T029 \\ 
    Biomedical Occupation or Discipline      & bmod  & T091 \\ 
    Biomedical or Dental Material            & bodm  & T122 \\ 
    Body Part, Organ, or Organ Component     & bpoc  & T023 \\ 
    Body Space or Junction                   & bsoj  & T030 \\ 
    Cell Component                           & celc  & T026 \\ 
    Cell Function                            & celf  & T043 \\ 
    Cell                                     & cell  & T025 \\ 
    Congenital Abnormality                   & cgab  & T019 \\ 
    Chemical                                 & chem  & T103 \\ 
    Chemical Viewed Functionally             & chvf  & T120 \\ 
    Chemical Viewed Structurally             & chvs  & T104 \\ 
    Clinical Attribute                       & clna  & T201 \\ 
    Clinical Drug                            & clnd  & T200 \\ 
    Cell or Molecular Dysfunction            & comd  & T049 \\ 
    Carbohydrate Sequence                    & crbs  & T088 \\ 
    Drug Delivery Device                     & drdd  & T203 \\ 
    Disease or Syndrome                      & dsyn  & T047 \\ 
    Environmental Effect of Humans           & eehu  & T069 \\ 
    Element, Ion, or Isotope                 & elii  & T196 \\ 
    Embryonic Structure                      & emst  & T018 \\ 
    Enzyme                                   & enzy  & T126 \\ 
    Eukaryote                                & euka  & T204 \\ 
    Fully Formed Anatomical Structure        & ffas  & T021 \\ 
    Fish                                     & fish  & T013 \\ 
    Finding                                  & fndg  & T033 \\ 
    Fungus                                   & fngs  & T004 \\ 
    Food                                     & food  & T168 \\ 
    Genetic Function                         & genf  & T045 \\ 
    Geographic Area                          & geoa  & T083 \\ 
    Gene or Genome                           & gngm  & T028 \\ 
    Group Attribute                          & grpa  & T102 \\ 
    Group                                    & grup  & T096 \\ 
    Human-caused Phenomenon or Process       & hcpp  & T068 \\ 
    Health Care Related Organization         & hcro  & T093 \\ 
    Hazardous or Poisonous Substance         & hops  & T131 \\ 
    Hormone                                  & horm  & T125 \\ 
    Human                                    & humn  & T016 \\ 
    Immunologic Factor                       & imft  & T129 \\ 
    \bottomrule
\end{supertabular}
\clearpage
\begin{supertabular}{@{} p{10cm} c c @{}}
    \midrule
    Individual Behavior                      & inbe  & T055 \\ 
    Inorganic Chemical                       & inch  & T197 \\ 
    Injury or Poisoning                      & inpo  & T037 \\ 
    Indicator, Reagent, or Diagnostic Aid    & irda  & T130 \\ 
    Language                                 & lang  & T171 \\ 
    Laboratory Procedure                     & lbpr  & T059 \\ 
    Laboratory or Test Result                & lbtr  & T034 \\ 
    Mammal                                   & mamm  & T015 \\ 
    Medical Device                           & medd  & T074 \\ 
    Manufactured Object                      & mnob  & T073 \\ 
    Mental or Behavioral Dysfunction         & mobd  & T048 \\ 
    Molecular Function                       & moft  & T044 \\ 
    Molecular Sequence                       & mosq  & T085 \\ 
    Neoplastic Process                       & neop  & T191 \\ 
    Nucleic Acid, Nucleoside, or Nucleotide  & nnon  & T114 \\ 
    Nucleotide Sequence                      & nusq  & T086 \\ 
    Organic Chemical                         & orch  & T109 \\ 
    Organism Attribute                       & orga  & T032 \\ 
    Organism Function                        & orgf  & T040 \\ 
    Organism                                 & orgm  & T001 \\ 
    Organization                             & orgt  & T092 \\ 
    Organ or Tissue Function                 & ortf  & T042 \\ 
    Pathologic Function                      & patf  & T046 \\ 
    Physical Object                          & phob  & T072 \\ 
    Physiologic Function                     & phsf  & T039 \\ 
    Pharmacologic Substance                  & phsu  & T121 \\ 
    Plant                                    & plnt  & T002 \\ 
    Patient or Disabled Group                & podg  & T101 \\ 
    Receptor                                 & rcpt  & T192 \\ 
    Reptile                                  & rept  & T014 \\ 
    Research Device                          & resd  & T075 \\ 
    Virus                                    & virs  & T005 \\ 
    Vitamin                                  & vita  & T127 \\ 
    Vertebrate                               & vtbt  & T010 \\ 
    \bottomrule
\end{supertabular}
\clearpage
\clearpage
\section{Discarded Terms Table}
\label{appendix:discarded_terms}
\begin{supertabular}{@{} p{10cm} c c @{}}
    \toprule
    \textbf{Description} & \textbf{Code} & \textbf{ID} \\ 
    \midrule

    Activity                               & acty  & T052 \\ 
    Classification                         & clas  & T185 \\ 
    Conceptual Entity                      & cnce  & T077 \\ 
    Diagnostic Procedure                   & diap  & T060 \\ 
    Daily or Recreational Activity         & dora  & T056 \\ 
    Educational Activity                   & edac  & T065 \\ 
    Experimental Model of Disease          & emod  & T050 \\ 
    Entity                                 & enty  & T071 \\ 
    Event                                  & evnt  & T051 \\ 
    Family Group                           & famg  & T099 \\ 
    Functional Concept                     & ftcn  & T169 \\ 
    Governmental or Regulatory Activity    & gora  & T064 \\ 
    Health Care Activity                   & hlca  & T058 \\ 
    Idea or Concept                        & idcn  & T078 \\ 
    Intellectual Product                   & inpr  & T170 \\ 
    Molecular Biology Research Technique   & mbrt  & T063 \\ 
    Machine Activity                       & mcha  & T066 \\ 
    Mental Process                         & menp  & T041 \\ 
    Natural Phenomenon or Process          & npop  & T070 \\ 
    Occupational Activity                  & ocac  & T057 \\ 
    Occupation or Discipline               & ocdi  & T090 \\ 
    Phenomenon or Process                  & phpr  & T067 \\ 
    Population Group                       & popg  & T098 \\ 
    Professional or Occupational Group     & prog  & T097 \\ 
    Professional Society                   & pros  & T094 \\ 
    Qualitative Concept                    & qlco  & T080 \\ 
    Quantitative Concept                   & qnco  & T081 \\ 
    Research Activity                      & resa  & T062 \\ 
    Regulation or Law                      & rnlw  & T089 \\ 
    Substance                              & sbst  & T167 \\ 
    Self-help or Relief Organization       & shro  & T095 \\ 
    Social Behavior                        & socb  & T054 \\ 
    Spatial Concept                        & spco  & T082 \\
    \bottomrule
\end{supertabular}

\end{appendices}
\end{document}